\documentclass[format=acmsmall,screen=true,review=false,authorversion=true,prodmode,acmcsur]{acmart} %

\acmJournal{CSUR}

\usepackage[ruled]{algorithm2e}

\usepackage[nolist,nohyperlinks]{acronym}
\usepackage{url}
\usepackage{array}
\usepackage{longtable} 
\usepackage{booktabs}

\SetAlFnt{\small}
\SetAlCapFnt{\small}
\SetAlCapNameFnt{\small}
\SetAlCapHSkip{0pt}
\IncMargin{-\parindent}

\acmVolume{51}
\acmNumber{1}
\acmArticle{8}
\acmYear{2018}
\acmMonth{1}

\setcopyright{acmcopyright}

\acmDOI{10.1145/3150224}

\newcommand{\tbl}[1]{\caption{#1}}
\newcommand{\beginsummary}{\begin{tabular}{m{1.8cm}m{4cm}m{6.8cm}}}
\newcommand{\summaryheader}{\multicolumn{1}{c}{\textbf{Aspect}} & \multicolumn{1}{c}{\textbf{References}} & \multicolumn{1}{c}{\textbf{Key Takeaways}}}

\makeatletter
\def\subsubsection{\@startsection{subsubsection}{3}{10pt}%
                                 {-.5\baselineskip \@plus -2\p@ \@minus -.2\p@}%
                 {3.5\p@}{\@subsubsecfont}}
\makeatother

\begin{document}

\title[HPC Cloud for Scientific and Business Applications]{HPC Cloud for Scientific and Business Applications:\\ Taxonomy, Vision, and Research Challenges}

\author{MARCO A. S. NETTO}
\affiliation{IBM Research}
\orcid{orcid.org/0000-0002-0369-5827}

\author{RODRIGO N. CALHEIROS}
\affiliation{Western Sydney University}

\author{EDUARDO R. RODRIGUES}
\affiliation{IBM Research}

\author{RENATO L. F. CUNHA}
\affiliation{IBM Research}
\orcid{orcid.org/0000-0002-3196-3008}

\author{RAJKUMAR BUYYA}
\affiliation{The University of Melbourne}

\begin{acronym}
\acro{API}{Application Programming Interface}
\acro{BoT}{Bag-of-Tasks}
\acro{EC2}{Elastic Compute Cloud}
\acro{GFLOPS}{Giga FLoating-point Operations Per Second}
\acro{GPFS}{General Parallel File System}
\acro{GPU}{Graphical Processing Unit}
\acro{HCI}{Human-Computer Interaction}
\acro{HPC}{High Performance Computing}
\acro{IaaS}{Infrastructure-as-a-Service}
\acro{QoS}{Quality of Service}
\acro{SaaS}{Software-as-a-Service}
\acro{SSD}{Solid-State Drive}
\acro{SLA}{Service Level Agreement}
\acro{VM}{Virtual Machine}
\end{acronym}

\begin{abstract}
High Performance Computing (HPC) clouds are becoming an alternative to on-premise clusters for executing scientific applications and business analytics services. Most research efforts in HPC cloud aim to understand the cost-benefit of moving resource-intensive applications from on-premise environments to public cloud platforms. Industry trends show hybrid environments are the natural path to get the best of the on-premise and cloud resources---steady (and sensitive) workloads can run on on-premise resources and peak demand can leverage remote resources in a pay-as-you-go manner. Nevertheless, there are plenty of questions to be answered in HPC cloud, which range from how to extract the best performance of an unknown underlying platform to what services are essential to make its usage easier. Moreover, the discussion on the right pricing and contractual models to fit small and large users is relevant for the sustainability of HPC clouds. This paper brings a survey and taxonomy of efforts in HPC cloud and a vision on what we believe is ahead of us, including a set of research challenges that, once tackled, can help advance businesses and scientific discoveries. This becomes particularly relevant due to the fast increasing wave of new HPC applications coming from big data and artificial intelligence.
\end{abstract}



\authorsaddresses{%
Author's addresses: Marco A. S. Netto, IBM Research; Rodrigo N. Calheiros, Western Sydney University; Eduardo R. Rodrigues, IBM Research; Renato L. F. Cunha, IBM Research; Rajkumar Buyya, The University of Melbourne.
}

\maketitle
\renewcommand{\shortauthors}{M. A. S. Netto, R. N. Calheiros, E. R. Rodrigues, R. L. F. Cunha, R. Buyya}

\section{Introduction}

In the early 90s, clusters of computers \cite{buyya1999high,sterling2002beowulf} became popular in High Performance Computing (HPC) environments due to their low cost compared to traditional supercomputers and mainframes. Computers with high processing power, fast network connections, and Linux were fundamental for this shift to occur. To this day, these clusters can handle complex computational problems in industries such as aerospace, life sciences, finance, and energy. They are managed by \emph{batch schedulers} \cite{feitelson1997theory} that receive user requests to run jobs, which are queued whenever resources are under heavy use. As Service Level Agreements (SLAs) are usually not in place in these environments, users have no visibility or concerns on costs of running jobs. However, large clusters do incur expenses and, when not properly managed, can generate resource wastage and poor quality of service.

Motivated by the different utilization levels of clusters around the globe and by the need to run even larger parallel programs, in the early 2000s, Grid Computing became relevant for the HPC community. Grids offer users access to powerful resources managed by autonomous administrative domains \cite{foster2003grid,foster2001anatomy}. The notion of monetary costs for running applications was soft, favoring a more collaborative model of resource sharing. Therefore, quality of service was not strict in Grids, having users relying on best-effort policies to run applications.

In the late 2000s, cloud computing \cite{armbrust2010view,mell2011nist,buyya2009cloud} was quickly increasing its maturity level and popularity, and studies started to emerge on the viability of executing HPC applications on remote cloud resources. These applications, which consume more resources than traditional cloud applications and usually are executed in batches rather than 24x7 services, range from parallel applications written in Message Passing Interface (MPI) \cite{gropp1996high,gropp1999using} to the newest big data \cite{reed2015exascale,assunccao2015big,bahrami2015role,dean2008mapreduce} and artificial intelligence applications---the latter mostly relying on deep learning \cite{coates2013deep,krizhevsky2012imagenet}. Cloud then came up as an evolution of a series of technologies, mainly on virtualization and computer networks, which facilitated both workload management and interaction with remote resources respectively. Apart from software and hardware, cloud offers a business model where users pay for resources on demand. Compared to traditional HPC environments, in clouds users can quickly adjust their resource pools, via a mechanism known as \emph{elasticity}, due to the size of the platforms managed by large cloud providers.

\textit{HPC cloud} refers to the use of cloud resources to run HPC applications. Parashar \textit{et al.}~\shortcite{parashar2013cloud} break down the usage of cloud for HPC into three categories: (i) ``HPC in the cloud'', which focuses on moving HPC applications to cloud environments; (ii) ``HPC plus cloud'', in which users use clouds to complement their HPC resources (a scenario known as cloud bursting to handle peak demands \cite{deassuncao2009evauating}); and (iii) ``HPC as a Service'', which exposes HPC resources via cloud services. These categories are related to how resources are allocated and abstractions to simplify the use of cloud.

HPC cloud still has various open issues. To exemplify, the abstraction of the underlying cloud infrastructure limits the tuning of HPC applications. Moreover, most cloud networks are not fast enough for large-scale tightly coupled applications---those with high inter-processor communication. The business model of HPC cloud is also an open field. Cloud providers stack several workloads on the same physical resources to explore economies of scale; an approach not always appropriate for HPC applications. In addition, although small companies benefit from fast access to resources from public clouds with no in advance notice, this is usually not true for large users. The market forces that operate at large scales of cloud computing are the same as for other products and services. If one wants large amounts of resources, other methods of delivery are more suitable, such as private clouds, customized long term contracts (e.g. Strategic Outsourcing), or even multi-party contracts.

Although several advances happened in the last years in the cloud space, there is still a lot to be done in HPC cloud. Studies have shown some challenges on HPC cloud \cite{vecchiola2009high,ubercloud2014,ubercloud2013,magellan2011cloud,mauch2013high,richter2016suitability,yang2014cloud,gantikow2015taxonomy,sterling2009high,galante2016analysis}, however, they do not present a comprehensive view of findings and challenges in the area. Therefore, this paper aims at helping users, research institutions, universities, and companies understand solutions in HPC cloud. These are organized via a taxonomy that considers the viability of HPC cloud, existing optimizations, and efforts to make this platform easier to be consumed. Moreover, we provide a vision with directions for the research community to tackle open challenges in this area.

\section{Taxonomy and Survey}
\label{sec:survey}

The main difficulties in using cloud to execute \ac{HPC} applications come from their properties in comparison to those from traditional cloud services such as standard enterprise and Web applications \cite{varghese2017next}. \ac{HPC} applications tend to require more computing power than cloud services. Such computing requirements come not only from CPUs, but also from the amount of memory and network speeds to support their proper execution. In addition, such applications have a particular execution mechanism compared to cloud services that run 24x7. HPC applications tend to be executed in batches. Users run a set of jobs, which are instances of the application with different inputs, and wait until results are generated to decide whether new jobs need to be executed. Therefore, moving \ac{HPC} applications to cloud platforms requires not only special care on resource allocation and optimizations in the infrastructure, but also on how users interact with this new environment. Therefore, proper understanding of all these aspects is necessary to bring HPC users to cloud platforms.

\begin{figure*}[!h]
        \centering
        \includegraphics[width=.75\linewidth]{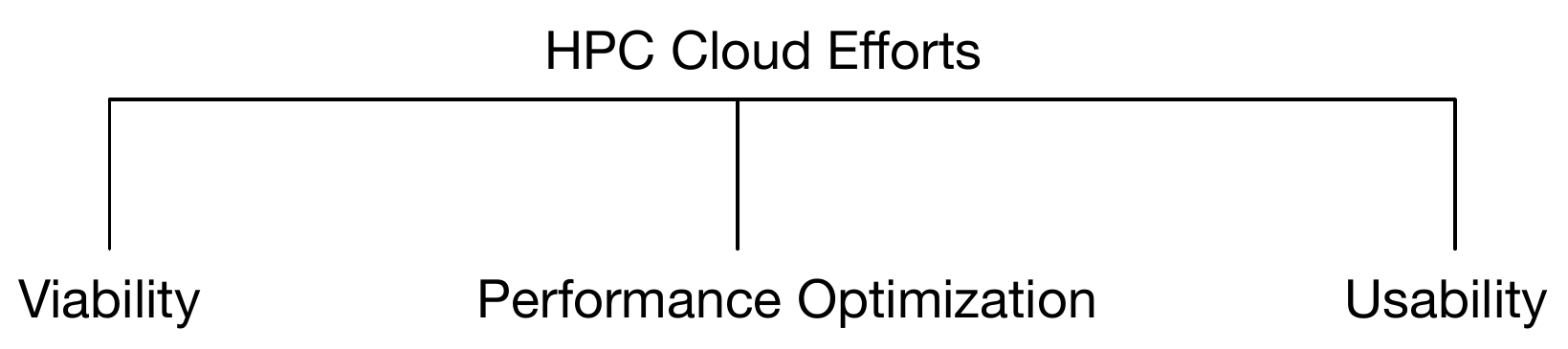}
        \caption{%
            Classification of HPC Cloud main research efforts.
        }\label{fig:main_tree}
\end{figure*}

Research in the area of HPC cloud can be classified in three broad categories, as depicted in Figure~\ref{fig:main_tree}: (i) viability studies on the use of cloud over on-premise clusters to execute HPC applications; (ii) performance optimization of cloud resources to execute HPC applications; and (iii) services to simplify the use of HPC cloud, in particular for non-IT specialized users.

For research in the first category, analyses were carried out by executing HPC benchmarks (mostly NPB---NAS Parallel Benchmark~\cite{bailey1991parallel}---and IMB---Intel MPI Benchmarks~\cite{intel2017imb}), microbenchmarks~\cite{barbosa2009Microbenchmark}, and HPC user applications; all focusing on CPU, memory, storage, and disk performance. A few studies utilized other types of applications such as scientific workflows~\cite{kwok1999WorkflowSurvey} and parameter sweep applications~\cite{casanova2000parameterSweep}. For cluster infrastructure comparisons, some studies utilized clusters interconnected via high bandwidth/low latency networks including the well-known Myrinet~\cite{boden1995Myrinet} and InfiniBand~\cite{ruivo2014exploring,vienne2012performance}, whereas other studies investigated performance of clusters interconnected via commodity Ethernet networks, with and without system virtualization.

Still in the first category, on the cloud side, the vast majority of the studies utilized Amazon Web Services (AWS)~\cite{aws}. This happened because AWS provided credits for researchers to utilize the cloud infrastructure and Amazon was the first player in the market which simplified the use of cloud resources for individuals and small organizations. More recent works compare different public cloud providers, an analysis that would not have been possible in the early days of cloud computing. Another effect of advances in the cloud technology over time in HPC research is the availability of HPC-optimized cloud resources~\cite{awshpc}. The first generation of such machines were made available by AWS in 2010, and thus earlier research did not investigate their performance. In this category, the Magellan report~\cite{magellan2011cloud} is the most comprehensive document in terms of range of architectures (i.e., number of different HPC clusters, virtualized clusters, and cloud providers), number of workloads (benchmarks and applications), and metrics. In fact, the report is a compilation of a number of studies sponsored by the U.S. Department of Energy (DOE) Office of Advanced Scientific Computing Research (ASCR) to study the viability of clouds to serve the Department's computing needs, which were at the time met mostly by on-premise HPC clusters. In addition, most of the viability studies are focused on public clouds because most large scale workloads from industry are not published in academic papers.

On the second category, \emph{i.e.}~on optimizing the performance of HPC clouds, targeted either the infrastructure level or the resource management level. In the former, networking has been the main target, as it was established that networking accounted for most of the inefficiencies of executing HPC workloads in the cloud. In the latter, scheduling policies that are aware of application and platform characteristics were proposed. In the optimization of resource allocation for HPC cloud, we observe a series of efforts on platform selectors due to hybrid cloud and multiple cloud choices, and studies aligned with specific features in cloud environments such as spot instances and elasticity. All of these optimizations benefit from resource usage and performance predictions.

Efforts in the third category focused on abstracting away the infrastructure from HPC users. One of the goals is to create ``HPC as a Service'' platforms where HPC applications are executed in the cloud without requiring users to have any understanding of the underlying cloud infrastructure. Users just submit the application, relevant parameters, and QoS expectations, such as deadline, via a web portal, and a middleware takes care of resource provisioning and application scheduling, deployment, and execution. This category of studies is relevant to increase the adoption of HPC, especially for new users with no expertise in system administration and configuration of complex computing environments. This category also highlights efforts on moving legacy applications to Software-as-a-Service deployments. This changes the user workflow from submitting jobs that wait in cluster scheduler queues to a cloud environment where resources are provisioned on demand according to user needs.

In the following sections we detail the work in each category, having a large body of the work on HPC cloud viability. After the survey we introduce our vision on what are the main missing components to enhance the capabilities of HPC cloud environments and their respective research challenges and opportunities.

\subsection{Viability: Performance and Cost Concerns}\label{sec:survey_viability}

\begin{figure*}[!h]
        \centering
        \includegraphics[width=1.0\linewidth]{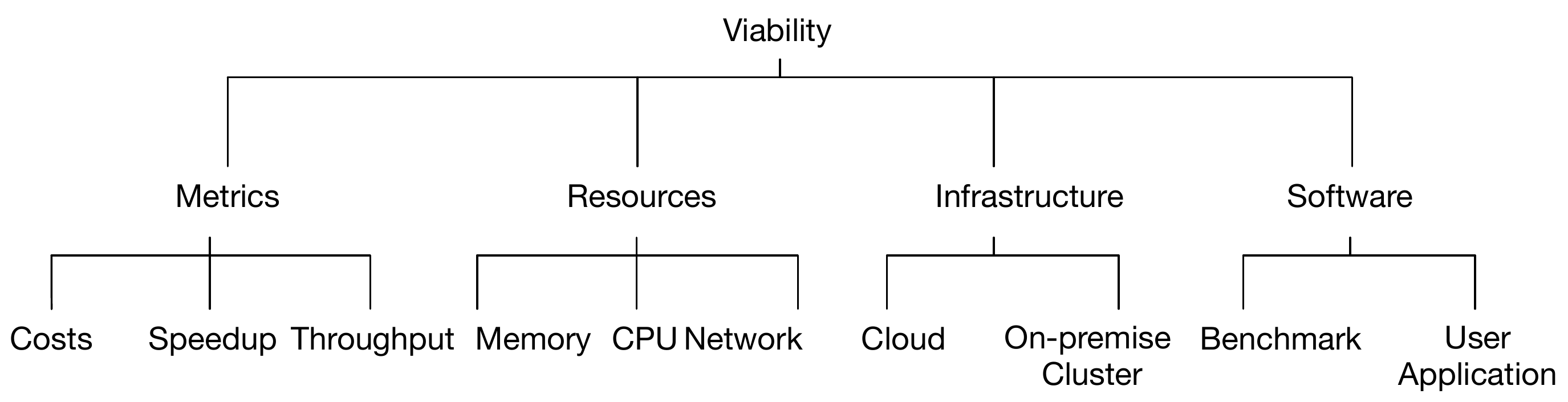}
        \caption{Classification of HPC Cloud viability studies.}
        \label{fig:viability_tree}
\end{figure*}

There are four main aspects that were considered in viability studies as depicted in Figure~\ref{fig:viability_tree}: (i) metrics used to evaluate how viable it is to use HPC cloud; (ii) resources used in the experiments; (iii) computational infrastructure; and (iv) software, which comprised well-known HPC benchmarks and user applications.

Gupta \textit{et al.}~\shortcite{gupta2013who} ran experiments using benchmarks and applications on various computing environments, including supercomputers and clouds, to answer the question ``why and who should choose cloud for \ac{HPC}, for what applications, and how should cloud be used for \ac{HPC}?''. They also considered thin \acp{VM}\footnote{Virtual Machines written specifically to run on top of hypervisors with the objective of reducing overhead.}, OS-level containers~\cite{soltesz2007containers} \cite{felter2015updated}, and hypervisor- and application-level CPU affinity~\cite{love2003cpuAffinity}. They concluded that public clouds are cost-effective for small scale applications but can complement supercomputers (\textit{i.e.} HPC plus cloud \cite{parashar2013cloud}) using cloud bursting and application-aware mapping. They also mentioned that network latency is a key limitation for scalability of applications in the cloud. For the experiments, based on their analyses, they found out that a cost ratio between two times and three times is a proper approximation to capture the differences between both cluster and cloud environments. This cost ratio reflects the shift from CAPEX to OPEX. The in-house HPC environment cost includes hardware acquisition, facility, power, cooling and maintenance \cite{eubank2003design}, that are not directly managed by the user in the cloud environment. However, IT support may still be present in the cloud, since most HPC applications are user specific. For further details, Kashef and Altmann~\shortcite{kashef2011cost} present a cost model for hybrid clouds itemizing all elements that go into the in-house and cloud environments.

Gupta and Milojicic \textit{et al.}~\shortcite{gupta2011evaluation} highlighted that cloud can be suitable for only a subset of \ac{HPC} applications; and for the same application, the choice of the environment may depend on the number of processors used. Gupta \textit{et al.}~\shortcite{gupta2012exploring} also remarked that obtaining application signatures (or characterization) is a challenging problem, but with substantial benefits in terms of cost savings and performance. They evaluated the performance of HPC benchmarks across clusters, grids, and a private cloud. The analysis confirmed that HPC applications can have performance degradation in clouds if they are communication-intensive, but they can achieve a good performance otherwise. Furthermore, if the cost of running an HPC infrastructure is taken into consideration, the cost-performance ratio can be in favor of the cloud for HPC applications that do not demand high-performance networks.

The study from Gupta \textit{et al.}~\shortcite{gupta2014evaluating} was further expanded for more platforms---including public cloud providers and public HPC-optimized clouds---and more applications. They identified different classes of applications, considering cloud scalability, driven by different communication patterns and the ratio between number of messages and message sizes. The cause for the differences in scalability has been identified as network virtualization, multi-tenancy, and hardware heterogeneity. Based on these findings, authors identified two general strategies for countering performance limitations of clouds called cloud-aware HPC and HPC-aware clouds. The first is about decomposing work units, setting up optimal problem size, and tuning network parameters to improve computation/communication ratio; and the second is about using lightweight virtualization, setting up CPU affinity, and handling network aggregation to reduce the overhead of the underlying virtualization platform. Their study also identified that, the more CPU cores an application requires, the more likely an HPC platform offers best value-for-money, although startups and small and medium size companies, which are usually sensitive to CAPEX, might still be better off using clouds rather than clusters.

Napper and Bientinesi~\shortcite{napper2009cancloud} used High-Performance LINPACK~\cite{dongarra2003Linpack} to evaluate whether cloud could potentially be included in the Top 500 list~\cite{top500}---the list of the most powerful computers worldwide. The experiments were conducted using Amazon \ac{EC2}. Their results showed that the raw performance of \ac{EC2} instances is compatible with resources from current on-premise systems. However, memory and network performance were not sufficient to scale the application. They also investigated \ac{GFLOPS} and GFLOP per dollar to evaluate the trade-off of costs and performance when running user applications in remote cloud resources. On the cloud, these metrics behaved differently from traditional \ac{HPC} systems. In November 2011, it was announced that an Amazon EC2 cluster reached position 42 in the Top 500 ranking.\footnote{\url{https://www.top500.org/list/2011/11/}}

Belgacem and Chopard~\shortcite{belgacem2015hybrid} investigated the use of hybrid \ac{HPC} clouds to run multi-scale large parallel applications. Their motivation was the lack of memory in their local cluster to run their application. They showed that using proper strategies to balance the load between local and remote sites had a considerable influence on the overall performance of this large parallel application, thus leading to the conclusion that \ac{HPC} hybrid clouds are relevant for large applications.

Marathe \textit{et al.}~\shortcite{marathe2013comparative} compared \ac{HPC} clusters against top-of-the-line \ac{EC2} clusters using two metrics: (i) turnaround time and (ii) total cost for executions. The turnaround time includes the expected queue wait time determined by the cluster management system, which is a commonly ignored but highly important factor for viability studies considering quality-of-service. They showed that although the clusters produced superior raw performance, \ac{EC2} was able to produce better turnaround times. The results showed turnaround times more than four times longer in \ac{HPC} on-premise resources than in the cloud, even with much faster executions on local clusters. They also highlighted that applications should be properly mapped to clusters, which are not generally the most powerful ones. And the choice between cloud and local \ac{HPC} clusters is complicated---which relies on application scalability and the goal of optimizing cost or turnaround time. Therefore, having tools that help in making resource allocation decisions is fundamental in order to save money and time.

Exp{\'o}sito \textit{et al.}~\shortcite{exposito2013performance} studied the performance of HPC applications in Amazon EC2 resources and focused mainly on I/O and scalability aspects. In particular, they compared CC1 instances against the CC2 instances released in 2011 and used up to 512 cores. They also investigated the cost-benefit of using these instances to run HPC applications. Their conclusions were that although CC2 provides more raw and point-to-point communication performance, collective-based communication-intensive applications performed worse compared to using CC1. They also concluded that using multi-level parallelism---one level of message passing and another with multithreading---generated a scalable and cost-effective alternative for user applications in Amazon EC2 CC instances. Such instances were also investigated by Sadooghi and Raicu~\shortcite{sadooghi2013understanding} using High-Performance LINPACK Benchmark and they found instabilities in network latency compared to their on premise HPC environment.

Carlyle \textit{et al.}~\shortcite{carlyle2010cost} conducted an experiment to identify if it would be worth, cost-wise, using Amazon EC2 cluster instances rather than their two clusters at Purdue University. The motivation of their study was the lack of cost-aware studies between community clusters and public clouds. Their conclusion was that community cluster had a better cost benefit than using cloud, especially given the high utilization of their clusters. However, they acknowledged that for low utilization clusters or small and underfunded projects, cloud could be a cost-effective alternative.

Ostermann \textit{et al.}~\shortcite{ostermann2009performance} presented a detailed study on the performance of EC2 for scientific computing. They evaluated EC2 instances using multiple benchmarks and considered various aspects including time to acquire and release resources, computing performance, I/O performance, memory hierarchy performance, and reliability. Although they found performance and reliability to be the major limitations for scientific computing, they found cloud as attractive for scientists in need of resources immediately and temporarily.

Zaspel and Griebel~\shortcite{zaspel2011massively} reported performance results of a parallel application in the area of computational fluid dynamics. They ran experiments on Amazon EC2 instances using both CPUs and GPUs to evaluate the application scalability. The application was executed with up to 256 CPU cores and 16 GPUs, and their findings indicated that the application scaled well until 64 CPU cores and 8 GPUs.

Berriman \textit{et al.}~\shortcite{berriman2010application} relied on three applications with workflow models to study the benefit of a cluster at the National Center for Supercomputing Applications (NCSA) against Amazon EC2 cloud offers. The workflows are Montage\footnote{\url{http://montage.ipac.caltech.edu}} from astronomy, Broadband\footnote{\url{http://scec.usc.edu/research/cme}} from seismology, and Epigenome\footnote{\url{http://epigenome.usc.edu}} from biochemistry. They concluded that, for their experiments, cloud can provide resources that are more powerful and cheaper compared to their on-premise cluster, in particular, considering applications that are CPU and memory intensive. However, for applications that manipulate large volumes of data, clusters with parallel file systems and low latency networks provide better performance. They also highlighted the growing efforts on creating academic clouds, which, according to them, may have different levels of services than those provided by commercial clouds. %

Zhai \textit{et al.}~\shortcite{zhai2011cloud} described in the SuperComputing'11 conference a detailed study comparing cloud and an in-house cluster considering tightly coupled applications. They investigated the EC2 cluster compute instances released by Amazon in 2010. They used NAS benchmarks and three parallel applications with different characteristics. Similar to the other studies they observed the network as a bottleneck for tightly coupled applications. However, they also evaluated which types of messages were not suitable for the 10GB Ethernet offered by these instances. They concluded that for large messages with few processors, the network performance was comparable to their in-house cluster. They also compared the costs of both environments with a detailed analysis on the required utilization level a cluster should have to be considered cheaper or more expensive than running HPC applications in the cloud. They observed that depending on the application, clusters should have at least 8.5\% to 31.1\% of utilization level to be beneficial compared to cloud---as highlighted by the authors, these numbers were calculated with a set of assumptions that brought benefit to the in-house cluster. 

The Magellan Report~\cite{magellan2011cloud}, commissioned by the U.S. Department of Energy (DoE), is one of the most comprehensive documents in the area of adoption of clouds for scientific HPC workloads. Extensive evaluation of diverse cloud architectures and systems, and comparison with HPC clusters, have been carried out as part of the Report's activities. Certain workloads from the DoE suffered slowdown of up to 50 times when utilized in clouds, due to the particular patterns of communication between application tasks. The report highlights that latency-limited applications (where numerous small point-to-point messages are exchanged) are the most penalized ones by the lack of high performance networks in clouds, whereas bandwidth-limited applications (exchanging few large messages, or performing collective communication) are less penalized. Another obstacle for HPC cloud, noted in the report, is the eventual absence of support, in the hypervisor level, to specialized instruction sets. When such instructions are enabled at the hypervisor, no performance loss is incurred during computation. Furthermore, when users cannot enforce a certain CPU architecture (as in the case in most public IaaS providers), CPU set-specific compiler optimizations cannot be utilized, what also limit the  potential increase in performance of applications. On the positive side for cloud adoption, the report notes that high performance cluster queueing systems do not provide proper support to embarrassingly parallel applications, and thus this class of applications can benefit from clouds.

Roloff \textit{et al.}~\shortcite{roloff2012high} described a study on HPC cloud considering three well-known cloud offers (Microsoft Azure, Amazon EC2, and Rackspace), and analyzed three aspects, deployment of HPC applications, performance, and costs, utilizing NAS benchmarks. A few highlights from their study are: (i) there is no single clear provider that best meets all three aspects analyzed; (ii) in various scenarios cloud showed an interesting alternative for on-premise cluster to run HPC applications considering both raw performance and monetary costs; and (iii) the lack of information on network interconnection among the cloud resources is still an issue for all cloud providers, in particular for communication-intensive application. Authors also described in details the deployment strategies offered by each provider.

Egwutuoha \textit{et al.}~\shortcite{egwutuoha2013cost} compared IaaS against not an on-premise cluster, but against a cloud provider that offers bare-metal machines, termed HaaS (Hardware-as-a-Service). By using HPL benchmark and an application called ClustalW-MPI for bioinformatics sequence alignment, they were able to show that HaaS can save up to 20\% the cost to run the applications compared to a traditional IaaS provider.

Evangelinos and Hill~\shortcite{evangelinos2008cloud} reported their experience using cloud to run their atmosphere-ocean climate model parallel application. Their experiment started by running NAS benchmarks to understand the cloud performance. Next, they tested various MPI implementations and highlighted the issue of not being able to control in which subnet the virtual machines were provisioned. Having the environment setup they tested their application and concluded that EC2 instances were a feasible alternative for their experiments, which considered only performance aspects.

He \textit{et al.}~\shortcite{he2010case} ran experiments to verify the performance of the NAS Parallel Benchmark, LINPACK, and an application on climate and numerical weather prediction application using three cloud environments: Amazon EC2 Cloud, GoGrid Cloud, and IBM Cloud. They concluded that with a few changes from a cloud perspective, mainly in relation to network and memory, cloud could be an alternative from on-premise clusters for HPC applications. They highlighted that different from traditional on-premise HPC platforms, in which FLOPS is the main optimization criterion, in cloud FLOPS per-dollar is an important metric. From their experiments, one of the examples was that changing an execution from one environment to another produced a gain of 30\% of performance but paying 4 times more. So, users can get to a point where they will accept having 30\% slower executions but with a set of resources that is 4 times cheaper.

Jackson \textit{et al.}~\shortcite{jackson2010performance} evaluated the performance of the NERSC benchmarking framework, which contains application internals from several fields including astrophysics, climate, and materials science. They compared Amazon EC2 against three on-premise clusters. They also used the Integrated Performance Monitoring framework \cite{borrill2005integrated} which helps determine the different application phases, \textit{i.e.} computing and communication. Their main conclusion is that cloud is suitable for various applications but not, in particular, for tightly-coupled ones. They found a considerable relationship between the amount of time an application spends in communication and its performance when using cloud resources. Similar findings were discussed by Ekanayake and Fox~\shortcite{ekanayake2009high} for MapReduce technologies, by Church and Goscinski~\shortcite{church2011iaas} for bioinformatics and physics applications, and by Hill and Humphrey~\shortcite{hill2009quantitative} with STREAM memory bandwidth benchmark and Intel's MPI Benchmark.

Hassan \textit{et al.}~\shortcite{hassan2015scalability} investigated the performance of Intel MPI Benchmark suite (IMB) and NAS Parallel Benchmarks (NPB) on Microsoft Azure on 16 virtual machines. They were more interested in analyzing scalability of these benchmarks considering different point-to-point communication approaches using both MPICH and OpenMPI. They obtained more promising results when running experiments on single virtual machines due to the shared memory communication as inter-node communication showed to be a key bottleneck.

Hassani \textit{et al.}~\shortcite{hassani2014improving} implemented a version of the Radix sorting algorithm using OpenMPI and tested it on both their on-premise cluster and Amazon EC2 extra large instances. They varied the input size of data to be sorted and concluded that the cloud environment generated 20\% faster completion time compared to their on-premise cluster using up to 8 nodes. After that point, they highlight that network bandwidth could limit the performance if instances were not reserved in advance.

\begin{table}[!t] %
\begingroup
\centering
    \tbl{Overview of the related work on viability of HPC cloud.\label{tab:viability_overview}}{%
    \renewcommand*{\arraystretch}{1.5}
        \beginsummary%
        \toprule
        \summaryheader\\
        \midrule
        Cost & \cite{gupta2013who,napper2009cancloud,carlyle2010cost,roloff2012high} &  Related to financially-related decisions. When compared with low utilization clusters and when running small applications, cloud environments can be preferred over on-premise clusters. \\
        Throughput & \cite{marathe2013comparative,carlyle2010cost,ostermann2009performance} &  Related to job turn around times. For running single applications with low communication requirements, cloud can have higher throughput due to the lack of queueing systems present in on-premise clusters. \\ 
        Resources & \cite{gupta2014evaluating,napper2009cancloud,exposito2013performance,sadooghi2013understanding,zhai2011cloud,egwutuoha2013cost,hassan2015scalability} &  Related to how different resource types impact performance of the applications. Network virtualization and hardware heterogeneity are main causes of poor performance of HPC in cloud; single-node performance is comparable between environments; improvements in infrastructure affect HPC applications positively.\\
        Network & \cite{magellan2011cloud,gupta2011evaluation,gupta2012exploring,napper2009cancloud,sadooghi2013understanding,zhai2011cloud,jackson2010performance,he2010case,evangelinos2008cloud} &  Related to the impact of network speeds considering different communication models of parallel applications. Cloud is suitable when loosely-coupled or embarrassingly-parallel applications are used.\\ 
        \bottomrule
    \end{tabular}}
\endgroup
\end{table} %

\begin{table}[!t] %
\begingroup
\centering
    \tbl{Recommendation to use cloud.\label{tab:viability_apptype}}{%
    \renewcommand*{\arraystretch}{1.5}
        \begin{tabular}{m{2.5cm}m{10.5cm}}
        \toprule
        \multicolumn{1}{c}{\textbf{Application type}} & \multicolumn{1}{c}{\textbf{Recommendation}}\\
        \midrule
        Large-scale tightly coupled & They are typical MPI applications that use thousands of cores and require high-performance network, such as weather, seismic, geomechanical and computational fluid dynamics models (time stepped applications). Any virtualization bottleneck and high latency network will have negative impact on application performance. Therefore, it is recommended to use these applications in traditional supercomputing centers, or on private clouds featuring baremetal machines and high-speed networks.\\
        Mid-range tightly coupled & They utilize a number of cores ranging from tens to hundreds and have lower performance requirements than the large-scale tightly coupled type. Consequently they are more tolerant of virtualization and traditional networks. Event driven simulation is an example of this type, but also other time stepped applications whose jobs are less deadline-sensitive. Its recommended that these applications explore the benefits of fast resource access to the cloud, especially when lightweight virtualization (i.e. containers) are becoming pervasive.\\
        High throughput & They are composed of independent tasks that require little or no communication, popular in Monte Carlo simulations and many other bag-of-tasks and map-reduce applications, and can benefit from variable number of available resources and are tolerant of resource heterogeneity. It is recommended the use of cloud for these applications especially by exploring elasticity mechanisms. They can also benefit from HPC hybrid cloud environments by spreading out tasks on both on-premise clusters and public clouds.\\
        \bottomrule
    \end{tabular}}
\endgroup
\end{table} %

\medskip
\noindent \textbf{Summary and takeaways.}  Overall, all these studies showed that cloud has a great potential to host HPC applications, with network performance as one of the main bottlenecks at the moment. Fortunately, this network bottleneck may not be a problem for several scientific and business applications that are CPU intensive. Most of the studies still rely on standard HPC benchmarks to stress different resource requirements of HPC applications. In addition, users and institutions should avoid limiting their cloud vs on-premise cluster decision based only on raw application performance. It is important to understand turn around time, that is execution time plus the time to access resources, costs, and resource demands. A mix of on-premise cluster and cloud seems to be a proper environment to balance this equation. The key takeaways of the papers referenced in this section are summarized in Table~\ref{tab:viability_overview}. We also provide in Table~\ref{tab:viability_apptype} a list of recommendations on when cloud could be used depending on application types defined in the Magellan report~\cite{magellan2011cloud}.

\subsection{Performance Optimization: Resource Allocation and Job Placement}\label{sec:survey_services}

\begin{figure*}[!h]
        \centering
        \includegraphics[width=1.0\linewidth]{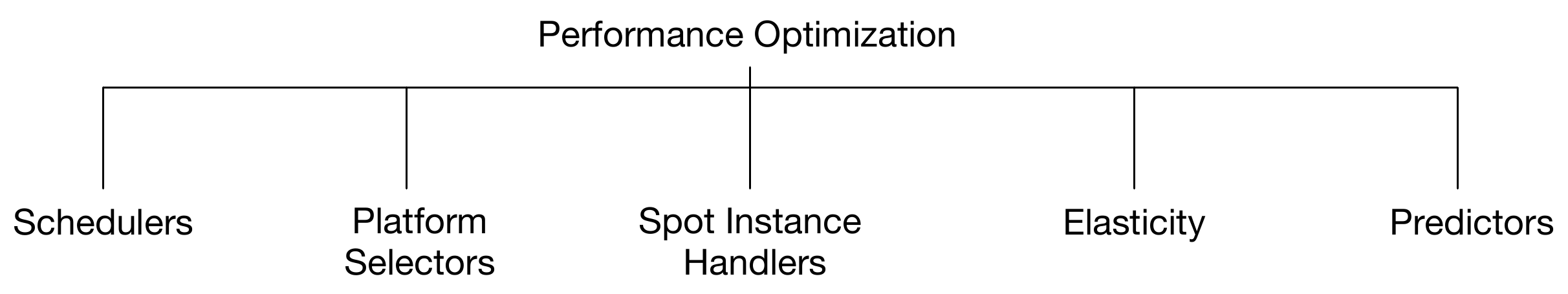}
        \caption{%
            Classification of HPC cloud performance optimization studies.%
        }\label{fig:services_tree}
\end{figure*}

As depicted in Figure~\ref{fig:services_tree}, most of the work on performance optimization for HPC cloud lies on areas related to resource management and job placement systems. These efforts concern where and how jobs should be placed, inside a cloud or in a hybrid environment (HPC in the cloud and HPC plus cloud according to Parashar \textit{et al.}~\shortcite{parashar2013cloud} definitions), how to leverage cheaper instances, how to use elasticity of cloud resources, and how to use prediction systems for job placement decisions.

Gupta \textit{et al.}~\shortcite{gupta2013hpc} introduced an HPC-aware scheduler for cloud platforms that considers topology requirements from HPC applications. Their scheduler utilizes benchmarking information, which classifies the type of network requirement of the application and how its performance is affected when resources are shared with other applications. Their experiments, using three applications and NAS benchmarks, showed performance improvements of up to 45\% by using their scheduler compared to an HPC-agnostic scheduler. In a related topic, Gupta \textit{et al.}~\shortcite{gupta2012exploring} developed a tool to extract characteristics from HPC applications and then map them to the most suitable computing platform, including both clusters and clouds.

Church \textit{et al.}~\shortcite{church2015exposing} addressed the issue of resource selection in the Uncinus framework. The selection is based on pre-populated information on available HPC applications and resources, which can be clouds and clusters. Resource selection is then carried out with the use of historical information on cloud resource usage and availability of resources at request time. The approach relies on a broker that mediates access to the cloud (via credentials, resource discovery, and resource selection) and enables sharing of resources and applications. Their goal is to help non-IT specialized users deploy their applications in the cloud.

Gupta \textit{et al.}~\shortcite{gupta2014evaluating} proposed a set of heuristics for the problem of choosing a platform (including clusters and clouds) for a stream of jobs. The objective is to improve makespan and job completion time. They tested a combination of static and dynamic heuristics that are application-aware and application-agnostic. Results demonstrated that dynamic heuristics that consider application characteristics and their performance on each particular platform generate better throughput than other heuristics.

Ashwini \textit{et al.}~\shortcite{ashwini2013efficient} introduced a framework to allocate cloud resources for HPC applications. Their motivation is the heterogeneity of physical cloud resources and their framework can select resources based on similar performance levels. They used processing power and point-to-point communication bandwidth to select clusters of VMs to run HPC applications.

Marathe \textit{et al.}~\shortcite{marathe2014exploiting} designed and implemented techniques to reduce costs when running HPC applications in the cloud. They developed techniques for both determining bid prices for Amazon EC2 spot instances, and scheduling checkpoints of user applications. They were able to obtain gains of 7x compared to traditional on-demand instances.

Somasundaram and Govindarajan~\shortcite{somasundaram2014cloudrb} developed a framework to schedule HPC applications in cloud platforms. Their goal was to meet users' deadline and at the same time reduce costs to run their applications. The scheduler is based on Particle Swarm Optimization and their experiments were based on simulations and on two real applications and a testbed setup with Eucalyptus Cloud middleware~\cite{nurmi2008Eucalyptus}.

Netto \textit{et al.}~\shortcite{netto2015deciding} discussed the challenges users have when making job placement decisions in HPC hybrid clouds, in particular with respect to uncertainties coming from execution time and job waiting time predictions. As a follow up of this problem, Cunha \textit{et al.}~\shortcite{cunha2017job} presented a detailed implementation of an advisor that relies on job run time and wait time predictions and the confidence level of such predictions to make job placement decisions. On the direction of using performance predictions, Shi \textit{et al.}~\shortcite{shi2012program} applied Amdahl's law to predict performance of NAS benchmarks over a private HPC cloud environment.

Network is another type of resource that requires proper management. Mauch  \textit{et al.}~\shortcite{mauch2013high} investigated the use and configuration of InfiniBand~\cite{infiniband2000infiniband} in virtualized environments. Their motivation is the limitation of the Ethernet technology used by cloud providers for HPC workloads. Their HPC cloud architecture based on InfiniBand allows customers to allocate virtual clusters isolated from other tenants in the physical network. Such platform, augmented with the capacity of establishing elastic \emph{virtual clusters} for users and achieving network isolation, was shown to incur only a small overhead in the order of microseconds. They also envisioned as future directions how to use InfiniBand to allow live migration of virtual machines.

Marshall \textit{et al.}~\shortcite{marshall2013high} provided an overview of their work on HPC cloud having elasticity as their main cloud functionality to be explored for HPC workloads. They developed a prototype using Torque and Amazon EC2 in which jobs would receive machines in the cloud if not enough resources were available at the job submission moment. They raised a series of questions they believe are relevant for HPC workloads related to when cloud machines should be provisioned, \textit{i.e.} if they should be provisioned at the moment new jobs arrive or once they stay stuck in queue, if instances should be deleted once jobs are completed or should remain until the hour completes as new jobs may arrive, and if auto-scaling should be done in a reactive or proactive manner.

Mateescu \textit{et al.}~\shortcite{mateescu2011hybrid} proposed a concept they called ``Elastic Cluster'', which aims at combining the strengths of clouds, grids, and clusters by offering intelligent infrastructure and workload management that are aware of different types and locations of resource and performance requirements of applications. Performance guarantees are obtained via a combination of statistical reservation and dynamic provisioning. The architecture is complemented by the capacity of setting \emph{personal virtual clusters} for execution of user workflows and by the capacity of combining resources from multiple Elastic Clusters for execution of workflows. Elasticy was also explored by Righi \textit{et al.}~\shortcite{righi2016autoelastic} who created a software at the platform-as-a-service to support this functionality for HPC applications. The aim of their project is to offer elasticity with no need of application source code access.

Zhang et al.~\shortcite{zhang2016high} studied the use of container-based virtualization and its relationship with the performance of MPI applications. Authors found out that the overhead on the computation stage of applications is negligible, although performance degradation can be observed in the communication stage, even when all the containers hosting application processes are executed on the same physical server. To tackle such an issue, the authors proposed a method to enable the MPI runtime to detect processes that share the same physical host and use in-memory communication among them. The approach, combined with optimization of configuration parameters of the MPI runtime, improved in 900\% and 86\% the performance of point-to-point and collective communications, respectively.

Fan \textit{et al.}~\shortcite{fan2014topology,fan2012topology} proposed a framework that considers application internal communication patterns to deploy scientific applications. This topology-aware deployment framework consists in clustering cloud machines that have communication affinity. Experiments were conducted using the PlanetLab~\cite{chun2003Planetlab} environment with 100 machines and relied on NAS Parallel Benchmarks (NPB) and Intel MPI benchmark (IMB). They were able to reduce execution times by up to 30-35\% compared to not considering application and topology and inter-machine communication performance.

\medskip
\noindent \textbf{Summary and takeaways.}  All efforts described in this section indicate that even with challenges in HPC cloud environments, optimization techniques and technologies have helped improve the performance of user HPC applications running in remote cloud resources. These optimizations influence allocation of different types of resources including CPUs, memory, and computing networks. Performance prediction plays a key role to help in resource allocation decisions to make the best use of cloud platforms. In addition, users have to be careful with application placement decisions, in particular trying to explore possibilities of using powerful single nodes whenever possible. Most of these efforts were motivated by network and virtualization issues from current cloud platforms. Even with new virtualization technologies and high performance networks, such optimizations will still bring benefits to users. Table~\ref{tab:services_overview} summarizes the efforts discussed in this section.

\begin{table}[!t] %
\begingroup
\centering
    \tbl{Overview of the related work on performance optimization for HPC cloud.\label{tab:services_overview}}{%
    \renewcommand*{\arraystretch}{1.5}
        \beginsummary%
        \toprule
        \summaryheader\\
        \midrule
        Scheduler & \cite{gupta2013hpc,ashwini2013efficient,somasundaram2014cloudrb,fan2012topology,fan2014topology} &  Related to how scheduling decisions impact job performance. HPC-aware schedulers indeed improve performance of HPC applications in cloud environments as they exploit both HPC application and infrastructure properties.\\
        Platform \hspace{2mm} Selectors & \cite{gupta2014evaluating,netto2015deciding,cunha2017job,church2015exposing} & Related to the impact of environment selection on job performance. Automated selection of execution environments eases transition to the cloud as users may be overloaded with many infrastructure configuration choices. \\
        \shortstack{Spot Instance\\\hspace{-6.5mm}Handlers} & \cite{marathe2014exploiting} & Related to the use of novel pricing models in cloud. Exploiting different cloud pricing models transparently reduces costs in the cloud because users have different resource consumption patterns.\\
        Elasticity & \cite{mateescu2011hybrid,mauch2013high,zhang2016high,marshall2013high,righi2016autoelastic} &  Related to dynamic allocation of resources. Cloud elasticity adds more flexibility to HPC applications; VMs placed on same host improve performance.\\
        Predictors & \cite{shi2012program,netto2015deciding,cunha2017job} & Related to collect information of future resource consumptions. Prediction of expected run time and wait time helps on job placement decisions due to a proper match of resource configuration and resource consumption. \\
        \bottomrule
    \end{tabular}}
\endgroup
\end{table} %

\subsection{Usability: User Interaction and High-Level Services}\label{sec:survey_usability}

\begin{figure*}[!h]
        \centering
        \includegraphics[width=1.0\linewidth]{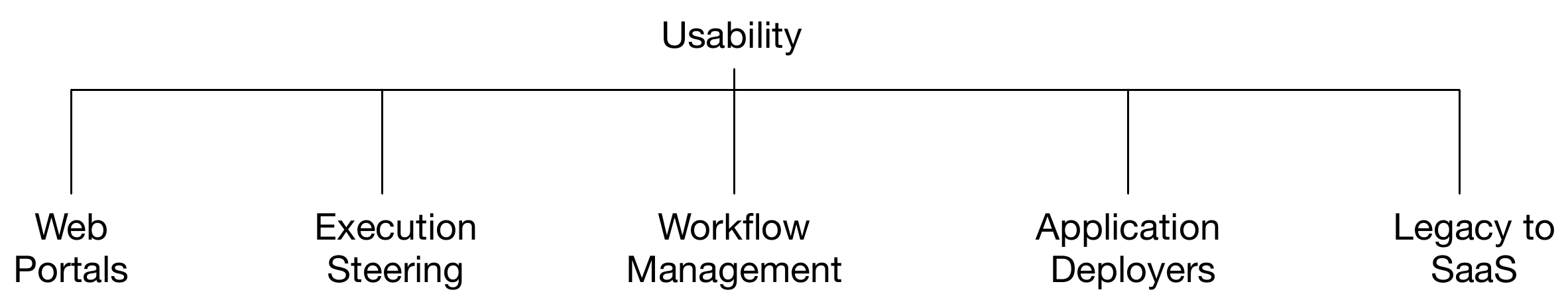}
        \caption{Classification of HPC Cloud usability efforts.}
        \label{fig:usability_tree}
\end{figure*}

Understanding the cost benefit of moving HPC workloads to the cloud and optimizing the execution of these workloads are the main efforts of researchers working in the area. However, all these efforts will have limited applicability if HPC cloud is not easy to use. Figure~\ref{fig:usability_tree} depicts the main research areas concerning the usability of HPC cloud. Most of the work relies on Web Portals and how users interact with their applications and the cloud infrastructure. There are also various efforts to support the creation of easy-to-use Software-as-a-Service based on legacy applications and offer HPC resources as a service (HPC as a service as defined by Parashar \textit{et al.}~\shortcite{parashar2013cloud}). 

Belgacem and Chopard~\shortcite{belgacem2015hybrid} reported their experience of porting a parallel 3D simulation in the computational fluid dynamics domain to a hybrid cloud consisting of a cluster located in Switzerland and AWS EC2 resources located in USA. To facilitate the integration between clusters and clouds, authors utilized a methodology called Distributed Multiscale Computing (DMC), which provides abstractions that enable phenomena to be described in different scales. Each scale becomes a parallel application (in the presented paper, written in MPI) to be executed in a physical computing resource.  These different scales can be processed in parallel or can hold dependencies among themselves, leading to a workflow model for application execution. Results demonstrated that utilization of a hybrid infrastructure only improves execution time if the application is tuned to adapt to the difference in CPU speeds between clusters and clouds and, more importantly, to adjust to the difference in speeds between local networks and WAN connections. This can be done, for example, by adapting the application to allow for overlap between communication and computation, thus reducing the time tasks are waiting for data to compute.

In the area of computational steering, SciPhy \cite{ocana2011sciphy} is a tool to help scientists in the area of Phylogeny/Phylogenomics run experiments for drug design. The tool, based on SciCumulus \cite{de2010scicumulus}, uses cloud resources and is an example on how a service can be created to facilitate the use of cloud for non-IT specialists. We believe several efforts on computational steering \cite{mattoso2015dynamic} applied to Grid and cluster computing will be leveraged by cloud users.

By using a middleware called Aneka, Vecchiola \textit{et al.}~\shortcite{vecchiola2009high} showed two case studies on HPC cloud: (i) gene expression data classification and (ii) fMRI brain imaging. Both applications were executed in Amazon EC2. They highlighted research directions on helping users in performance/cost trade-offs and decision-making tools to specify accuracy level of the experiment or parts of data to be processed according to predefined Service Level Agreements.

Huang~\shortcite{huang2014development} created a mineral physics SaaS, called Fonon, for on-premise HPC clusters. Usage of Fonon allowed users of different backgrounds to submit jobs more easily and to obtain more useful results. The benefits achieved come from users submitting and analyzing jobs by means of a web-based application. From his analysis, although he noticed the virtualization overhead and network limitations when experimenting the application in a public cloud, he also identified a demand to facilitate the use of the application through the creation of a service. Therefore, his SaaS encapsulated several activities of the users such as execution of pre-processing and post-processing scripts, elimination of data movement by creating figures and tables in the cloud provider site, and allocation of cluster resources. In similar direction, Buyya and Diana~\shortcite{buyya2015multi} described the use of Aneka to help users run and deploy HPC applications on multi-cloud environments. They used BLAST as an example of a parameter sweeping application that could benefit from Aneka application deployment system.

In 2012, Church \textit{et al.}~\shortcite{church2012toward} discussed the difficulties users, especially those with no IT expertise, had to execute applications using remote cloud resources. Having this motivation, they started to develop a framework to translate HPC applications into services. Over the years, the same group \cite{church2015exposing} presented novel use cases of their framework in the area of genomics. Church and Goscinki~\shortcite{church2014survey} also presented a survey on technologies available to help researchers working with mammalian genomics run their experiments. They highlighted the technical difficulties these researchers have when using IaaS. They then developed a system to encapsulate several activities related to resource allocation, data management, images containing software stacks, and graphical user interfaces.

Abdelbaky \textit{et al.}~\shortcite{abdelbaky2012enabling} described a system to enable HPCaaS using an application that models oil reservoir flows as a use case. Users are able to easily interact with the application, which has access to IBM BlueGene/P systems and can provide dynamic resource allocation (\textit{e.g.}\ elasticity). The system relies on both DeepCloud and CometCloud which provide IaaS and PaaS functionalities respectively. Also in the context of CometCloud, Abdelbaky \textit{et al.}~\shortcite{abdelbaky2014exploring} proposed a Software-as-a-Service abstraction for scientists to run experiments using distributed resources; as use case they considered an application for experimental chemists to use mobile devices to run Dissipative Particle Dynamics experiments.

Wong and Goscinski~\shortcite{wong2013unified} created a software to configure, deploy, and offer HPC applications in the cloud as services. The approach also includes mechanisms for discovery of already deployed applications---thus avoiding duplication of the endeavor. Once deployed, the application becomes available for end users via a web-based portal, and the only input required in this case are input parameters, which are supplied via web forms. Petcu et al.~\cite{petcu2014next} proposed a different framework with similar objectives, with the added difference of including support to hardware accelerators, such as GPUs.

Balis \textit{et al.}~\shortcite{balis2016porting} presented a methodology for porting HPC applications to a cloud environment using a multi-frontal solver as a case study. The methodology focuses on a task agglomeration heuristic to increase task granularity while ensuring there is enough memory to run them, a task scheduler to increase data locality, and a two-level storage to enable in-memory storage of intermediate data.

\begin{table}[!t] %
\begingroup
\centering
    \tbl{Overview of the related work on usability of HPC cloud.\label{tab:usability_overview}}{%
    \renewcommand*{\arraystretch}{1.5}
        \beginsummary%
        \toprule
        \summaryheader\\
        \midrule
        Web Portals & \cite{huang2014development,church2012toward,church2014survey,church2015exposing,abdelbaky2012enabling,wong2013unified,petcu2014next} & Related to creation of easy-to-use interfaces. Users have a hard time selecting cloud resources; developing portals that abstract details can increase user productivity. \\
        Execution Steering & \cite{ocana2011sciphy,de2010scicumulus} &  Related to automatic creation of jobs with different input parameters. Facilitation of parameter sweeping experiments for non-IT specialists. \\
        Workflow Management & \cite{vecchiola2009high,belgacem2015hybrid,church2015exposing,church2014survey,wong2013unified} & Related to proper mapping of multiple activities, possibly including dependencies. Frameworks that have knowledge of cloud pricing models can reduce time and costs for deploying and exposing HPC applications in the cloud. \\
        Application Deployers & \cite{wong2013unified,bunch2011neptune,buyya2015multi} & Related to HPC application software stack dependencies. HPC applications have a different software stack than traditional cloud applications; systems that are aware of HPC application requirements reduce time to solution in clouds. \\
        Legacy-to-SaaS & \cite{balis2016porting,petcu2014next,huang2014development} & Related to transform legacy applications running in traditional computing platforms to cloud environments. Direct ports of legacy applications to cloud environments might have inefficiencies due to different infrastructure assumptions; principled methodologies can overcome such inefficiencies. \\
        \bottomrule
    \end{tabular}}
\endgroup
\end{table} %

Another effort to simplify the usage of HPC comes from Bunch \textit{et al.}~\shortcite{bunch2011neptune}, who created a domain specific language, called Neptune, to deploy HPC applications in the cloud. Neptune offers support to applications written with various packages, including MPI, X10, and Hadoop. It can also be used to add and remove resources to the underlying cloud platform and to control how applications are placed across multiple cloud infrastructures.

\medskip
\noindent \textbf{Summary and takeaways.} As described in Table~\ref{tab:usability_overview}, there has been a growing interest in creating services to facilitate the use of cloud for HPC applications and to transform legacy applications into cloud services. HPC is still not easy to be used by non-IT experts and having cloud services, even if in private environments, can help HPC become more popular. With these services, elasticity, which is a key functionality of cloud, could be embedded and explored more easily by end-users. There is a considerable engineering effort to improve usability of HPC cloud, such as the creation of Web portals or simplification of procedures to allocate and access remote resources. However, research can also make contributions. For instance, when transforming an HPC application into SaaS, the amount of computing resources to meet user expected QoS needs to be properly defined. This may also be an opportunity of collaboration with researchers working with human computer interface. In the area of workload management, languages for non-IT experts could also be beneficial to facilitate the use of cloud resources. Several of these efforts could leverage the work done in grid computing, however in HPC cloud, cost management is a crucial aspect. Another relevant aspect of usability is to increase user productivity. It may be more valuable for several users to reduce their time setup an experiment than optimizing resources to run jobs. The value of proper usability technologies and practices is to reduce turn around times and minimize costs whenever possible.

\section{Vision and Research Challenges}

As presented in the previous section, plenty of work has been done in HPC cloud. However, clients and cloud providers can benefit more from this platform if additional modules/functionalities become available. Here we discuss a vision and research challenges for HPC cloud and relate them to what was presented in the previous section.

\begin{figure*}[!b]
        \centering
        \includegraphics[width=.95\linewidth]{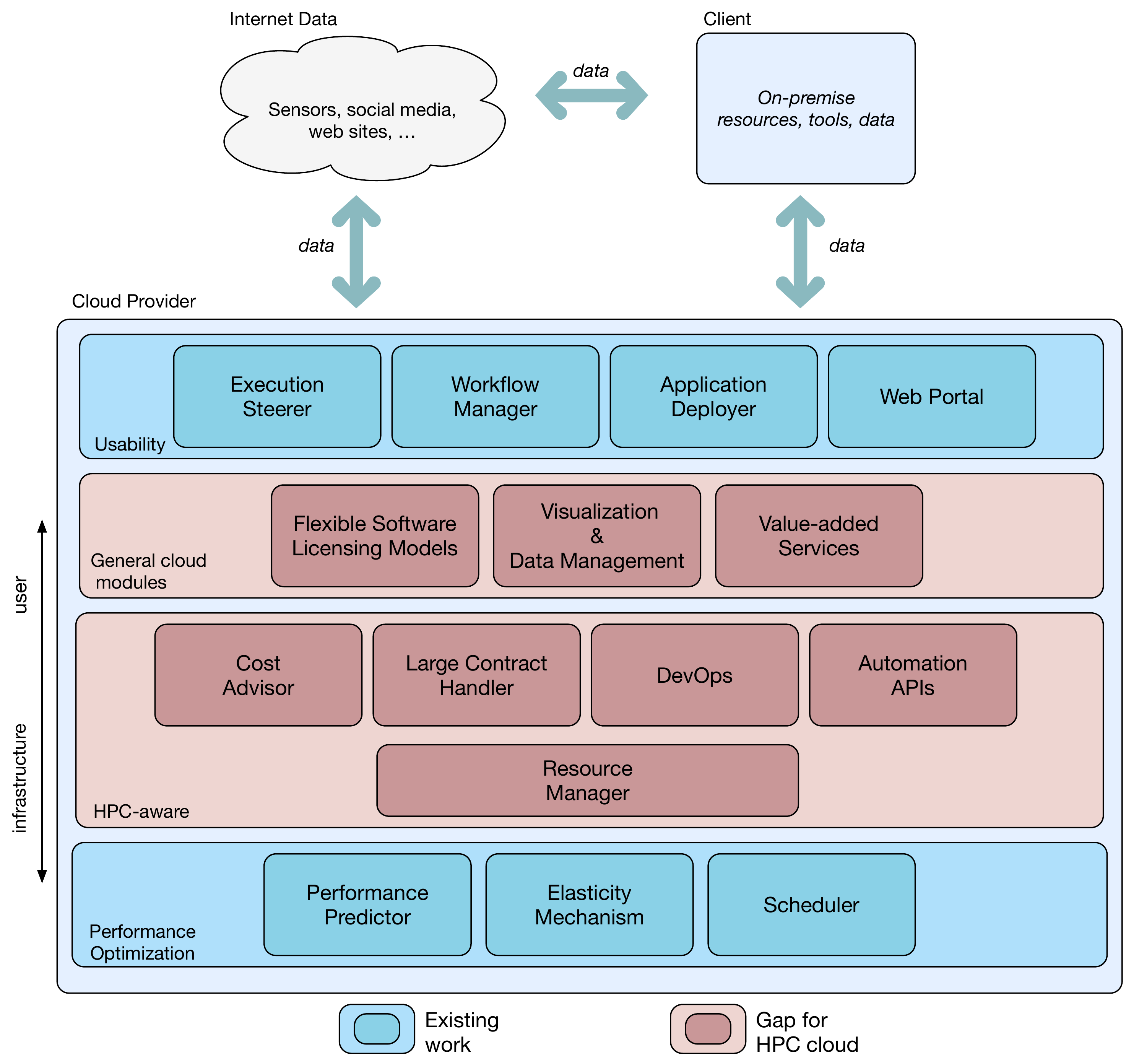}
        \caption{%
            Vision of an HPC cloud architecture comprising existing modules in the area of usability and performance optimization and modules that require further development. The latter modules have two categories: (i) HPC-aware modules and (ii) general cloud modules that bring benefit to HPC cloud.
        }\label{fig:arch_vision}
\end{figure*}

Figure~\ref{fig:arch_vision} illustrates our vision of an HPC cloud architecture. The architecture contains three major players: the Internet, the client, and the cloud provider. Sensors and social media networks are two relevant sources of data generation that serve as input for various HPC applications, especially for the increasing workloads coming from big data, artificial intelligence, and sensor-based stream computing. Data from these sources can come from places that are both internal and external to client and cloud provider environments. The other two players are the client who needs data to be processed and the cloud provider, who offers HPC-aware services. The architecture comprises components already discussed in Section~\ref{sec:survey}, including those from performance optimization, which are more related to infrastructure, and those from HPC cloud usability, which are closer to the user. In the following sections we discuss a set of modules that we believe require further development and we split them in two categories; one specific for HPC workloads and the other general for cloud but that can bring benefits to HPC cloud.

\subsection{HPC-aware Modules}

In this section we discuss challenges and research opportunities for five modules that require further development to meet the needs of HPC users. Although most of these modules are already available in various cloud providers, HPC users have different resource requirements and work style that limit the direct utilization of these modules.

\subsubsection{Resource Manager}

Cloud and HPC environments have distinct ways to manage computing resources. Cloud aims at consolidating applications into the same hardware to achieve economies of scale, which is possible due to virtualization technologies. To extract the best performance of a cloud environment, ongoing efforts have focused on increasing inter-VM isolation and reducing VM overhead. HPC environments, on the other hand, aim at bringing the highest performance possible from the infrastructure. User requests to access exclusively portions of a cluster are queued whenever resources are overloaded.  One evident benefit of cloud is exactly that no queues are required due to the ``unlimited'' availability of resources. In addition, HPC hardware, especially network, costs considerably more than those used to build traditional clouds. Hence, placing users in such hardware would be wasteful for cloud providers. Therefore, the challenge on HPC cloud resource management is to have sustainable business for cloud providers via economies of scale and be able to offer users high performance. Research efforts could find this balance and use the same manager for both cloud and HPC users.

There are a few projects already in this area. For instance, Kocoloski \textit{et al.}~\shortcite{kocoloski2012case} introduced a system called dual stack virtualization which consists of a VM manager that can host both HPC and commodity VMs depending on the level of isolation required by the users. The configurable level of isolation can relate to different prices that benefit both users and cloud providers.

To advance this area, we envision cloud providers offering queues and having pricing models that consider how long users are willing to wait to access HPC resources. This would allow providers to have more clusters fine-tuned for certain types of workloads and for users to have different QoS with respect to the time to access resources. There are some efforts on providing flexible resource rental models for HPC cloud, such as those from Zhao and Li~\shortcite{zhao2012designing}, who were motivated by the different interests from multiple parties when allocating resources. Their models are based on planning strategies that consider on-demand and spot instances. Resource managers can also have user-centric policies~\cite{sherwani2004libra} for management of shared HPC cloud resources that offer incentive for users to reveal true QoS requirements, such as deadlines to finalize application executions.

\subsubsection{Cost Advisor}

Cost advisor has become popular in several cloud providers. It is usually implemented as a simulator for users to specify their resource requirements and obtain cost estimations. Different from traditional cloud users who use this advisor to plan for hosting a service, HPC users need advice on how much their experiments will cost. Therefore, current cost advisors need to be adapted to support HPC users, and this is challenging because HPC user workflows involve tuning their applications and exploring scenarios via execution of several jobs. Such cost comes from software licenses and powerful computing resources which tend to be much more expensive than traditional lightweight virtual machines.

Researchers have been looking into solutions to handle cost predictions for HPC users. For instance, Aversa \textit{et al.}~\shortcite{aversa2011performance} highlighted that cost is a critical issue for HPC cloud because applications are not optimized for an unknown and virtualized environment and current charging mechanisms do not take into account the peculiarities of HPC users. They also noticed that such an advisory system for cost-related decisions is highly dependent on the user profile, which can vary from a typical HPC user who wonders the execution time for her job and the configuration of resources to obtain the highest performance/cost ratio to a high-end user who cares more on performance.

Rak \textit{et al.}~\shortcite{rak2015early} introduced an approach to help users have better predictions of performance and costs when running HPC in the cloud. Using their framework called mOSAIC, users can run simulations and benchmarks to give insights about application performance during the application development phase. Another example is the work from Li \textit{et al.}~\shortcite{li2011building} who investigated the problem of how to enhance interactivity for HPC cloud services. Their motivation is that HPC users have complex and expensive requirements. Therefore, for them, HPC cloud services are not only about helping users allocate resources, but also their interactions with the computing environment, which include their expectations. Their work allows users to predict how much longer a job will take to complete. Such information can help users reconfigure jobs, and authors showed that users can reduce costs with proper selection of resource configurations as the application reaches 10-20\% of completion.

The Cost Advisor module needs not only to be able to predict how long the jobs of a user will take to run, but also, how long an experiment composed of unknown jobs will take to run. HPC users usually run experiments in batches where they submit a group of jobs, analyze the produced results, and create new jobs based on the intermediate findings. Understanding this workflow, and giving feedback to user on estimations of costs is crucial to see if the ongoing strategy of running experiments can meet budget restrictions~\cite{silva2016sla}. Advisors need also to consider data storage and movement, which is common for HPC users.

\subsubsection{Large Contract Handler}

Most of the work presented in the survey comes from academic papers, which reflects one community of HPC\@. Enterprises with
large HPC demands also look for alternatives to run their compute-intensive applications. For these enterprises, current HPC pricing models may not be sustainable depending on their current HPC infrastructure utilization levels. If the utilization is high, it might be more beneficial for them to maintain their own clusters, but if they use their clusters for sporadic projects, cloud becomes a cost-beneficial alternative. For those enterprises with high cluster utilizations, cloud providers need to come up with sustainable models that are beneficial for them and for their clients. This is a challenging work and a rich opportunity for research projects as it involves capacity planning, theory for sustainable business models, negotiation protocols for multiple parties, and admission control mechanisms.

For large users, such as enterprises, a contract model for the cloud might be more appropriate. This model works similarly to what is currently seen in other
markets, such as energy: instead of buying energy on the spot, large users make contracts with electricity providers, for example. The objective of
these contracts is to minimize risks, such as fluctuations in price and discontinuities in supply. In some cases, the contract model for clouds may
impose some limitations in elasticity. A contract may specify a minimum and maximum size or amount of resources. For companies which have more
predictable workloads, this can still be suitable, since, by outsourcing infrastructure management, they can focus on their businesses.
Even though there are few publications in the scientific literature in this aspect, there are Requests for Proposals (RFPs) that show some of these trends \cite{gao2013ibmorder,noaa2016}.

HPC environments are different from a traditional cloud infrastructure---clusters tend to have jobs with static number of resources whereas cloud has as attractive the support for elastic jobs and shared resources. It would be expensive, and a waste, to set up clusters with InfiniBand for users that do not run HPC jobs. Therefore, clusters with such high speed networks need to be well sized because they are not easily expanded/shrunk compared to traditional cloud environments. It is then crucial for the cloud provider to have proper estimates of the demand to use such clusters. In addition, relying on a single client to rent a cluster may have negative impact if the cluster is not rented at full capacity or, even worse, if the client gives up on using the cluster.

One possible sustainable model is a multi-party contract, where multiple parties can have a contract with a cloud provider to create a managed infrastructure that can meet the demand of a group of clients. This helps clients have reduced costs and a cloud provider to set up an HPC environment that is suitable and easier to be managed and reduces risks. In this contractual model, cloud providers could offer partitions of large clusters with high speed networks they can offer to multiple clients. Whenever clients ask for more resources, depending on the overall demand of all or part of the clients, the cloud provider can increase the cluster size.

\subsubsection{DevOps}

DevOps aims at integrating development (Dev) and operations (Ops) efforts to enable faster software delivery~\cite{huttermann2012devops}. DevOps has become popular with the maturity of cloud computing as an important mechanism for both development and hosting of services. In the HPC world, where most applications are still built to run on on-premise clusters, there is still several opportunities to create tools to support DevOps for HPC workloads and platforms---the challenge is that HPC workloads are resource intensive and tests can become financially prohibitive. A few projects are pursuing the development of such technologies. For instance, the work from Rak \textit{et al.}~\shortcite{rak2015early}, presented in the previous section, helps developers have insights on application performance, which is an important component of DevOps for the HPC community.

Tests of HPC applications can be much more complex and resource consuming than traditional web applications being hosted in clouds. The fact that HPC software developers try their best to develop applications that not only work, but are optimized to run in parallel using several resources, the development workflow can become slow. In addition, as cloud resources are usually not exclusive, the performance of the application under development can vary each time the developer runs a test. Therefore, there is a great opportunity for researchers working with DevOps for HPC to create a set of services to facilitate software development. HPC users would benefit from research studies to help understand the balance between different types of tests (those with lighter or heavier resource consumers) considering the various possible performance levels cloud can offer. DevOps for HPC could also facilitate the use of elasticity \cite{righi2016autoelastic}, which is a key differentiator of cloud computing compared to traditional cluster environments. As DevOps explores the concept of tests, in an HPC cloud environment, it could also consider different prices for using computing resources depending on estimations of the type of tests necessary to run and the tolerable resource noise from other users.

\subsubsection{Automation APIs}

In spite of many efforts to create GUIs that allow drag-and-drop of components for software development and execution, automation is a cultural aspect that needs to be considered for the HPC community. The more automation the more comfortable HPC users will be with a cloud platform. Examples of activities that require automation APIs are: running a software system with hundreds or thousands of different input parameters; automating when a job should be executed in the cloud or on-premise; defining when new resources should be allocated or existing ones should be released are examples of activities that require automation APIs.

Moreover, several HPC users have job submission scripts that were refined over the years. Scientific and business users also have scripts that handle data input and output. Such scripts need to be leveraged so users do not start in the cloud from scratch. The challenge in the area of automation APIs is to be able to reuse existing scripts from traditional HPC environments and be able to easily extend them to explore the peculiarities and benefits of cloud, such as elasticity and management of resource allocation as a function of available budget the users have to run jobs. Researchers with background in software engineering can play a key role on how to achieve a great level of simplicity on this process. Researchers with background on human-computer interface can perform studies to understand which mechanisms of automation increase the productivity of users in this platform.

\subsection{General Cloud Modules}

In this section we discuss challenges and research opportunities for three general cloud modules that bring benefits to HPC cloud environments.

\subsubsection{Visualization and Data Management}

Usually data management and visualization are handled separately and we claim here they need to be more integrated, especially considering HPC and big data. It is well-known that data movement between on-premise and cloud infrastructure is a road blocker for several users~\cite{gantikow2015taxonomy}. Therefore, it is essential for a cloud provider to offer a service that helps users determine which data needs to be moved from one place to another. It is common for HPC users to process data in batches, with analysis and planning happening between batches. Visualization and data management, when integrated, can minimize data movement, allowing users to analyze intermediate results via remote visualization. Then, when ultimately needed, data can be transferred. This is a challenging area because it involves detection and modeling of user experience, predictors related to time and costs, and proper visualizations that bring value to users.

To create this module, multiple software components need to be in place. One of them is an estimator to help users determine the amount of time to transfer data between cloud and on-premise environments. Another component needs to be created to determine if a user is inside a session, that is, the user is submitting a stream of jobs with regular think times between new batches of jobs~\cite{zakay2012identifying}. By determining if the user is in a session, it is possible to provide proper suggestions on transferring data vs visualizing it remotely. The determination of the think time can also serve as a clue to understand if most of the time between the submission of new jobs is consumed by the user data analysis or the data transfer itself.

It is also relevant to enable users to verify the progress of their executions with rich visualizations. Users could monitor different metrics at both system level and application level. These visualizations may help users identify if it is worth downloading data or even continuing their ongoing executions, which would then save precious time and money. Some of these visualizations could be done by using intermediate output files generated by applications. If users are working with popular applications, cloud providers could offer them such visualizations as a service in their platforms.

\subsubsection{Flexible Software Licensing Models}

For several industries, HPC software licenses are expensive. As pointed out in the UberCloud reports~\cite{ubercloud2014,ubercloud2013} users still face unpredictable bills especially as they lack precise information on how long they need to run their applications and therefore for how long they will need to use software licenses.

There are several types of software licenses, including pay-per-use, shared license, floating license, site license, end-user license, and multi-feature license. Most existing HPC software systems offer licenses to enable their usage on on-premise user environments, which can be single servers or computer clusters. Over time, software companies are enabling cloud pay-per-use licenses.

We claim here that having more flexible software licensing models can help users and cloud and software providers reduce costs and increase usability respectively. Usage of a software system may not be flat nor peaky for several users. It may happen for periods of a few months and for a few hours a day, and therefore pay-per-use licenses depending on the prices may not be cost-effective for the users, which leads them to look for alternative software systems. A cloud provider, with possible partnerships of software companies, could offer flexible software licenses depending on the usage profile of the users, which could be pre-defined, or learned over time. Apart from bringing costs down to users, such flexibility may help users better determine their license expenses in advance, which is critical in HPC settings. In addition, this flexibility allows software companies to have broader usage of their software and cloud providers to attract more users to their environments. Multi-party, involving multiple clients through a marketplace, could also help reduce costs and increase usage of software as a service. Apart from a cultural change required by several companies to start exploring more the usage models of cloud, an interesting research area is to monitor access to software licenses and bring hints to the software owners on the value of different license models per client. This involves analytics and software consumption predictions. 

\subsubsection{Value-added Cloud Services}

As cloud computing matured, providers started building services on top of infrastructure resources, building higher value services for users. One example of value-added cloud services is the addition of Machine Learning
APIs to major providers' clouds, such as IBM Bluemix, Google Cloud Platform, and
Microsoft Azure. Most of the services currently available in these clouds are
meant for users to consume already-existing APIs of pre-trained models. Some
providers allow users to upload user-trained models for making predictions
using, for example, Tensorflow~\cite{abadi2016tensorflow} trained models. Current trends suggest the need to execute machine learning algorithms will increase in the coming years and, although deep learning methods have achieved very good performance in various domains, training procedures still lack in scalability when using Stochastic Gradient Descent (SGD)~\cite{keuper2016distributed,bhardwaj2016practical}, the default optimization method for neural networks. To improve the scalability of SGD-based learning algorithms, work has to be done in areas such as reducing communication overhead, parallelization of matrix operations and effective data organization and distribution, problems which the HPC community has experience solving. Improving the scalability of such algorithms would allow for better resource usage and larger HPC cloud clusters for machine learning~\cite{awan2017scaffe}.

As we observed there are several research directions in the area of HPC cloud. In this paper we highlighted ongoing research and in this section we described what we believe needs further work from the research community. Table~\ref{tab:challenges} summarizes these challenges.

\begin{table}[!t] %
\centering
    \tbl{Summary of research challenges to enhance HPC cloud capabilities.\label{tab:challenges}}{%
        \begin{tabular}{m{2.6cm}m{4.6cm}m{5.0cm}}
        \toprule
        \multicolumn{1}{c}{\textbf{Module}} & \multicolumn{1}{c}{\textbf{Importance}} & \multicolumn{1}{c}{\textbf{Research}}\\
        \midrule
        & \multicolumn{1}{c}{\textit{HPC-aware Modules}} & \\
        \cline{2-2} \\
        
        Resource Manager & Improve cloud performance for HPC workloads, sustainable HPC cloud business. & Handle HPC and cloud workloads under same management, new pricing models based on queue, support for queues in cloud.\smallskip\\

        Cost Advisor & Avoid unexpected costs for users. & Understand user workflow, predict future jobs and their performance.\smallskip\\
        
        Large Contract Handler & Sustainable HPC cloud business for cloud providers and clients. & Technology to allow simplified contracts for large users, facilitation of multi-party contracts.\smallskip\\

        DevOps & Bring DevOps benefits to HPC users. & Handle variable performance, minimize costs with resources, predict performance for various environments.\smallskip\\

        Automation APIs & Keep HPC user tooling in cloud environment. & Simplify migration of HPC scripts to cloud and easily add cloud functionalities to them.\smallskip\\

        & \multicolumn{1}{c}{\textit{General Cloud Modules}} & \\
        \cline{2-2} \\

        Visualization and Data Management & Improve user experience and reduce storage and data transfer costs.\smallskip & Identify user workflow, predict data transfer costs.\smallskip\\
        Flexible Software Licensing Models & Reduce costs for providers and clients.  & Software licensing models based on user workflows, multi-party licenses.\smallskip\\
        Value-added Cloud Services & Easy-to-use services for application development. & Encapsulate and optimize complex services to accelerate development.\smallskip\\
        \bottomrule
    \end{tabular}}
\end{table} %

\section{Concluding Remarks}

This paper introduced a taxonomy and survey for the existing efforts in HPC cloud and a vision architecture to expand HPC cloud adoption with its respective research opportunities and challenges. In the last years, cloud technologies became more mature, being able thus to support not only those initial e-commerce applications, but also more complex traditional HPC, big data, and artificial intelligence applications.

The attempts of moving applications with heavy CPU and memory requirements to the cloud started by verifying the cost-benefit of running those applications in the cloud against running on already owned on-premise clusters. Various researchers used well-known HPC benchmarks and a few applications also common in the area. The goal was to understand not only performance, but also monetary costs and how sustainable it would be to decommission their own clusters and move everything to the cloud. The main conclusion was that applications that were compute-intensive and with high inter-processor communication could not scale well in the cloud, especially due to the lack of low latency networks such as InfiniBand. However, a strong support seemed to be present when talking about embarrassingly parallel applications, which showed good performance with current cloud resources. There was also a visible concern about the difference of performance for multiple executions using the same group of allocated resources, which comes due to the resource sharing aspect of cloud computing.

Meanwhile, several other efforts started to emerge. Researchers started to question the time to provision new machines in the cloud, how cloud could host services to help researchers with no IT background, and how to properly allocate resources in the cloud, with a great focus on hybrid cloud. From a business point of view, hybrid clouds seem to be the current model that brings sustainability for several companies with HPC workloads. With this model, it is possible to leverage existing computing infrastructure and depending on the peak demands, part of workloads can be moved temporarily to the cloud. The amount of workload that should be moved to the cloud is highly dependent on the actual usage of existing resources---if utilization level is low, a strategy to reduce fixed capacity and use cloud for peak demands can become a cost-effective alternative.

With the increase of microservices, and technologies for DevOps, transforming existing HPC applications into Software-as-a-Service which are able to abstract the infrastructure layers can become a trend to make HPC more popular. In addition, research efforts can drive a better understanding of sustainable resource allocation pricing models for both cloud providers and HPC users. From a resource management perspective, it is well-known that, in several environments, users have the feeling of having access to free resources, making them allocate resources without assessing their actual needs. With cloud bringing the monetary aspect, user behavior may change in HPC settings, which also calls for research arms to have a better understanding about this shift.

As the in-house and cloud environments evolve, new technologies will appear. In this paper we have focused on what has been done in HPC cloud area. However, much research is devoted
to technologies that are not essentially HPC cloud, but that will eventually make to this environment. For example, new virtualization and containers technologies \cite{pahl2017cloud,zhang2016high,xavier2013performance} have been evolving and
will play a role to reduce the performance gap between on-premise clusters and public clouds. Moreover, fast and low latency networks will become more common\footnote{ProfitBricks Network: \url{https://www.profitbricks.com/cloud-networks}}\footnote{Azure Network: \url{https://azure.microsoft.com/en-us/pricing/details/cloud-services/}} \cite{zahid2016realizing,zhang2017designing,zhang2016high} and may reshape the current offerings found in the cloud.
New accelerators, such as GPUs \cite{giunta2010gpgpu,jermain2016gpu,li2017multimedia}, FPGAs \cite{iordache2016high,kachris2016survey}, TPUs \cite{jouppi2017datacenter}, and frameworks will become more pervasive and may make viable applications that currently are not present in the cloud environment.

Our main goal was to introduce a much broader perspective of interesting challenges and opportunities that HPC cloud can bring to researchers and practitioners. This is particularly relevant as HPC cloud platforms can become essential for new engineering and scientific discoveries, especially as HPC community starts to embrace new workloads coming from big data and artificial intelligence.

\section*{Acknowledgments}

We thank the anonymous reviewers for their helpful comments in the preparation of this article. This work has been partially supported by \textsc{FINEP/MCTI} under grant no. 03.14.0062.00.

\bibliographystyle{ACM-Reference-Format}
\bibliography{references}

\end{document}